\documentclass[twocolumn]{aastex631}

\shorttitle{Spectroscopic Orbits of Subsystems. X}

\begin{document}

\renewcommand{\topfraction}{1.0}
\renewcommand{\bottomfraction}{1.0}
\renewcommand{\textfraction}{0.0}

\newcommand{\kms}{km~s$^{-1}$\,}
\newcommand{\masyr}{mas~s$^{-1}$\,}
\newcommand{\msun}{$M_\odot$\,}

\title{Spectroscopic Orbits of Subsystems in  Multiple
  Stars. X (Summary)}

\author{Andrei Tokovinin}
\affiliation{Cerro Tololo Inter-American Observatory | NSF's NOIRLab
Casilla 603, La Serena, Chile}
\email{andrei.tokovinin@noirlab.edu}

\begin{abstract}
Results  of a  large  program of  spectroscopic  monitoring of  nearby
solar-type  stellar  hierarchical  systems using  the  CHIRON  echelle
spectrograph at  the 1.5  m telescope are  summarized.  Ten  papers of
this  series  contain  102   spectroscopic  orbits  and  substantially
contribute to  the knowledge  of periods and  eccentricties, providing
input for  the study of  their formation and early  evolution.  Radial
velocities of additional 91 targets  without CHIRON orbits (members of
wide physical pairs) are published  here.  Our results are compared to
the recent Gaia Non-Single Star (NSS) catalog, revealing its strengths
and weaknesses.   The NSS provides  orbital periods for 31  objects of
the  CHIRON sample  (about one  third).  Of  the 22  spectroscopic NSS
orbits in common, 14 are in  good agreement with CHIRON, the rest have
reduced  velocity amplitudes  or other  problems.  Hence  ground-based
monitoring  gives, so  far, a  more accurate  and complete  picture of
nearby hierarchies  than Gaia.  The  distribution of inner  periods in
hierarchical systems  is non-monotonic,  showing a shallow  minimum in
the 30-100 days bin and a strong excess at shorter periods, compared to
the  smooth  distribution  of  simple  binaries  in  the  field.   The
period-eccentricity  diagram  of  inner  subsystems  updated  by  this
survey,  recent   literature,  and   Gaia,  displays   an  interesting
structure.
\end{abstract}



\section{Introduction}
\label{sec:intro}

Observations of  spectroscopic subsystems  in nearby  solar-type stars
are motivated by the desire  to complement statistics of hierachies in
the  solar neighborhood  \citep{FG67b}.  In  most cases,  discovery of
such  subsystems  by  variable  radial velocity  (RV)  or  astrometric
acceleration has  not been  followed by  determination of  the orbits.
Without   knowledge   of   periods  and   mass   ratios,   statistical
distributions remain  poorly constrained,  hence useless as  input for
testing    models    of    formation   and    early    evolution    of
hierarchies. Development of predictive  models of stellar multiplitity
remains the ultimate goal that justifies new observations.

Monitoring of RVs is a classical way to find periods, mass ratios, and
orbital  eccentricities.  Such  long-term program  has been  conducted
since 2015  at the  1.5 m  telescope at Cerro  Tololo with  the CHIRON
high-resolution optical echelle spectrograph \citep{chiron}.  Its main
targets were solar-type stars  within 67\,pc belonging to hierarchical
systems  with  three  or  more   components.   The  program  has  been
complemented by hierarchies at  larger distances, also with solar-type
components.  Short-period  orbits could  be determined  rapidly, while
longer periods required observations  for several years. The resulting
orbits accompanied by discussions of  each hierarchy were published in
a series of 10 papers  listed in Table~\ref{tab:papers}, with a total
of  102 spectroscopic  orbits determined  throughout this  program.  A
summary of this effort is provided here.

The classical approach of monitoring  selected objects from the ground
is  nowadays   complemented  by   the  large   spectroscopic  surveys,
e.g. GALAH  \citep{GALAH} and LAMOST  \citep{LAMOST}, and by  the Gaia
space mission \citep{gaia3}. The Gaia Data Release 3 (GDR3) includes a
catalog  of non-single  stars (NSS)  which contains  about $3  \times 10^5$
spectroscopic    and/or    astrometric   orbits    \citep{Arenou2022}.
Uniformity   of  the   Gaia  coverage,   compared  to   the  selective
object-by-object  study,  is  a  huge advantage  for  the  statistics.
However, the current NSS suffers from incompleteness and selection and
contains a  non-negligible fraction of  wrong orbits, as noted  by its
compilers  \citep{Pourbaix2022}.  I  compare here  the NSS  and CHIRON
orbits to  highlight the  advantages and  caveats of  both approaches.
Taken  together,  the  ground- and  space-based  orbits  substantially
complement   our   knowledge   of  stellar   hierarchies   and   their
architecture.  Some  statistical results  are presented here  based on
the current version of the Multiple Star Catalog, MSC \citep{MSC} that
can         be         accessed        online.\footnote{         {\url
    http://www.ctio.noirlab.edu/\~{}atokovin/stars/}               and
  {\url http://vizier.u-strasbg.fr/viz-bin/VizieR-4?-source=J/ApJS/235/6}}

\begin{deluxetable}{c c r  r }
\tabletypesize{\scriptsize}     
\tablecaption{Publications on the CHIRON Survey
\label{tab:papers} }  
\tablewidth{0pt}                                   
\tablehead{                                                                     
\colhead{Paper\tablenotemark{a}} & 
\colhead{Bibcode} &
\colhead{$N_{\rm sys}$} & 
\colhead{$N_{\rm orb}$}  
}
\startdata
1  & 2016AJ....152...11T  & 4   &      7 \\
2  & 2016AJ....152...10T  & 4   &      7 \\
3  & 2018AJ....156...48T  & 6   &      9 \\
4  & 2018AJ....156..194T  & 9   &     10 \\
5  & 2019AJ....157...91T  & 9   &      9 \\
6  & 2019AJ....158..222T  & 11  &     12 \\
6a & 2020AJ....159...88T  & 1   &      2 \\
7  & 2020AJ....160...69T  & 8   &     12 \\
8  & 2022AJ....163..161T  & 10  &     19 \\  
9  & Submitted            & 14  &     15 \\
\enddata
\tablenotetext{a}{References: 
1 -- \citet{chiron1}; 
2 -- \citet{chiron2}; 
3 -- \citet{chiron3}; 
4 -- \citet{chiron4}; 
5 -- \citet{chiron5}; 
6 -- \citet{chiron6}; 
6a -- \citet{chiron6a}; 
7 -- \citet{chiron7}; 
8 -- \citet{chiron8}.
}
\end{deluxetable}

The  CHIRON multiplicity  project is  part of  a larger  observational
effort.  Studies  of stellar hierarchies  on the northern sky  using a
correlation  radial-velocity meter  were  conducted in  the 1990s,  as
summarized by  \citet{TS2002}, and continued in  the following decades
in     a     series     of     papers    by     Gorynya     et     al.
\citep{Gorynya2001,Gorynya2007,Gorynya2014,Gorynya2018}.  In parallel,
high-resolution imaging of wider (mostly astrometric) subsystems using
adaptive   optics   and    speckle   interferometry   was   undertaken
\citep{Tok2010,Tok2012};  it  is  continued   at  present.   With  all
techniques (including Gaia) combined, a  wide and deep coverage of the
full parameter space can be achieved for nearby stars.

The  CHIRON sample  is presented  in Section~\ref{sec:chiron}  and the
orbits are  compared to the  Gaia orbits in  Section~\ref{sec:dr3}. In
Section~\ref{sec:wide},  RVs  of  other targets  (mostly  wide  visual
binaries)  measured with  CHIRON are  published for  future use.   The
period distribution  and the  $P-e$ diagram  are discussed  briefly in
Section~\ref{sec:stat},    and    the     summary    is    given    in
Section~\ref{sec:sum}.

\section{Overview of the CHIRON program}
\label{sec:chiron}

The main 67-pc sample of hierarchies with solar-type primary stars was
based on the  Hipparcos catalog \citep{FG67a}, so it  is convenient to
use here  HIP numbers  as primary identifiers.   I extracted  from the
database of CHIRON  observations all Hipparcos stars  relevant to this
project,  omitting  a   few  systems  which  are   not  in  Hipparcos.
Table~\ref{tab:sample}  lists  120  individual  targets  (resolved  or
blended components  of multiple  systems) featured here.   The columns
contain the HIP  number, MSC/WDS code based on  the J2000 coordinates,
component  identifier,  its  equatorial  coordinates  for  J2000,  $V$
magnitude,  parallax  $\varpi$, its  source,  the  orbital period  $P$
determined  in  this  project,  and the  orbit  reference  (the  paper
number).   In some  cases the  components  of a  multiple system  have
individual  HIP numbers  (e.g.   HIP 6868  and  6873); otherwise,  the
secondary stars are  identified here by the HIP number  of the primary
and  a  component  letter  (e.g.    HIP  24320  A  and  B).   Detailed
information on  all components (accurate coordinates,  proper motions,
etc.)   can  be found  in  the  MSC.   As  indicated in  the  parallax
reference  column, most  parallaxes  come  from the  GDR3  or its  NSS
extension (DR3N).   If parallax  of the component  is not  measured in
GDR3, parallax of other system's components is used instead (DR3*). If
there are no wide components,  the parallax comes from Hipparcos (HIP)
or visual  orbits (dyn and  orb).  The  median parallax of  the CHIRON
sample is  17.3\,mas, 85 targets  are within 67\,pc  (parallaxes above
15\,mas).   Four targets  are revealed  as spectroscopic  triples with
inner periods of a  few days and outer periods on the  order of a year
(HIP 11537A, 27970A, 56282A,  111598A); both spectroscopic periods are
listed for those stars, which  also have outer visual companions (they
are rare quadruples of 3+1 hierarchy).

Candidates  for  determination  of spectroscopic  orbits  were  mostly
identified in the  survey by \citet{N04} and in  other publications as
components of visual binaries with variable RVs.  They are featured in
the original 67-pc  sample as hierarchies with  unknown inner periods.
Some  of   these  stars  also  have   astrometric  accelerations  (the
astrometric and spectroscopic binaries overlap).

Measurements of  the RVs of  wide (resolved) nearby binaries  with the
fiber echelle and CHIRON spectrographs at the CTIO 1.5 m telescope are
reported in  \citep{Tok2015}.  The  aim was  to detect  new subsystems
(one measurement of a substantial RV difference between the components
is sufficient  for a  detection).  This mini-survey  of 96  wide pairs
revealed 17  new subsystems which  were added to the  present program.
Components  of additional  wide multiples  were episodically  observed
with  CHIRON in  the  following  years as  a  complement  to the  main
program;  their  RVs  are  reported  here  in  Section~\ref{sec:wide}.
Typically, components of wide pairs were observed only once.

\startlongtable

\begin{deluxetable*}{r l l cc ccl c c c}    
\tabletypesize{\scriptsize}     
\tablecaption{Main CHIRON Sample
\label{tab:sample}          }
\tablewidth{0pt}                                   
\tablehead{                                                                     
\colhead{HIP} & 
\colhead{WDS} & 
\colhead{Comp.} & 
\colhead{R.A.} & 
\colhead{Decl.} & 
\colhead{$V$} & 
\colhead{$\varpi$} & 
\colhead{Ref.plx\tablenotemark{a}} & 
\colhead{$P$ }  &
\colhead{SB\tablenotemark{b} }  &
\colhead{Ref.}  \\
&  
\colhead{(J2000)} & &
\colhead{(deg)} & 
\colhead{(deg)} & 
\colhead{(mag)} & 
\colhead{(mas)} & &
\colhead{(days)} & &
}
\startdata
1103 & 00138+0812 & A & 3.440421 & 8.193534 & 7.40 & 13.19 & DR3 & \ldots & 0 & - \\
2863 & 00363-3818 & A & 9.073152 & -38.294137 & 8.36 & 9.56 & DR2 & 4.81 & 2 & 5 \\
3150 & 00400+1016 & A & 10.009878 & 10.266985 & 8.71 & 5.99 & DR3 & 172.5 & 2 & 7 \\
3645 & 00467-0426 & A & 11.668998 & -4.427128 & 7.58 & 30.08 & DR3 & 1530.9 & 2 & 9 \\
4974 & 01037-3024 & B & 15.932681 & -30.398708 & 8.78 & 6.40 & DR3 & 14.71 & 2 & 5 \\
6868 & 01284+0758 & A & 22.095514 & 7.961353 & 6.21 & 8.47 & DR3 & \ldots & 0 & 7 \\
6873 & 01284+0758 & B & 22.114598 & 7.958176 & 8.04 & 8.13 & DR3 & 115.6 & 2 & 7 \\
7601 & 01379-8259 & A & 24.481486 & -82.974996 & 5.89 & 37.01 & DR3 & 19.37 & 2 & 1 \\
7852 & 01410-0524 & A & 25.244875 & -5.403616 & 8.48 & 19.44 & DR3 & 1177.5 & 1 & 8 \\
8353 & 01477-4358 & A & 26.918957 & -43.971085 & 7.97 & 5.80 & HIP & 5.31 & 1 & 5 \\
9148 & 01579-2851 & A & 29.472168 & -28.846089 & 8.27 & 11.24 & DR3N & 1272.3 & 2 & 8 \\
9642 & 02039-4525 & A & 30.980214 & -45.412905 & 7.30 & 19.99 & DR3 & 4.78 & 2 & 2 \\
10710 & 02179+1941 & A & 34.481469 & 19.680560 & 8.91 & 15.17 & DR3 & \ldots & 0 & - \\
11537 & 02287-3114 & Aa & 37.182878 & -31.227254 & 9.45 & 15.52 & DR3 & 22.26 & 2 & 7 \\
11537 & 02287-3114 & Ac & 37.182878 & -31.227254 & 9.45 & 15.52 & DR3 & 1146.4 & 1 & 7 \\
12548 & 02415-7128 & A & 40.364023 & -71.462435 & 7.82 & 18.60 & HIP & 1851.8 & 1 & 8 \\
12548 & 02415-7128 & B & 40.364023 & -71.462435 & 7.82 & 18.60 & HIP & 108.2 & 1 & 8 \\
12780 & 02442-2530 & A & 41.060839 & -25.495420 & 6.96 & 21.68 & DR3 & 2443.2 & 2 & 2 \\
12779 & 02442-2530 & B & 41.060087 & -25.498774 & 8.50 & 22.44 & DR3 & 27.77 & 1 & 2 \\
13498 & 02539-4436 & A & 43.463080 & -44.605663 & 7.72 & 14.80 & HIP & 14.09 & 2 & 1 \\
14194 & 03030-0205 & B & 45.757457 & -2.086469 & 7.51 & 17.59 & DR3 & 1136.0 & 2 & 6 \\
14313 & 03046-5119 & B & 46.154110 & -51.318520 & 8.59 & 18.26 & orb & 6626.0 & 2 & 9 \\
16853 & 03369-4957 & A & 54.222556 & -49.958015 & 7.62 & 23.58 & DR3N & \ldots & 0 & - \\
19639 & 04125-3609 & A & 63.128760 & -36.152552 & 7.12 & 8.01 & DR3 & 2.35 & 2 & 4 \\
20375 & 04218-2146 & A & 65.447861 & -21.770803 & 7.54 & 18.99 & DR3 & \ldots & 0 & - \\
20752 & 04268-0143 & A & 66.703396 & -1.724609 & 8.03 & 20.13 & DR3 & \ldots & 0 & - \\
21079 & 04311-4522 & A & 67.769943 & -45.360427 & 8.29 & 21.87 & DR3 & 217.0 & 1 & 8 \\
22562 & 04509-5328 & C & 72.835632 & -53.405879 & 9.01 & 26.39 & DR3 & \ldots & 0 & 7 \\
22531 & 04509-5328 & A & 72.730398 & -53.461546 & 5.23 & 26.35 & DR3N & 1003.0 & 1 & 7 \\
22534 & 04509-5328 & B & 72.735411 & -53.459694 & 6.19 & 26.24 & DR3N & 208.3 & 2 & 7 \\
23824 & 05073-8352 & A & 76.827211 & -83.859952 & 6.80 & 21.16 & DR3 & 5.63 & 2 & 1 \\
24320 & 05131-5959 & A & 78.284645 & -59.982988 & 8.87 & 16.30 & DR3N & 1430.3 & 1 & 8 \\
24320 & 05131-5959 & B & 78.286273 & -59.984014 & 10.13 & 16.34 & DR3 & 7.94 & 1 & 8 \\
26444 & 05376+0607 & A & 84.404518 & 6.112005 & 7.65 & 20.18 & DR3 & \ldots & 0 & - \\
27970 & 05550-1256 & Aa & 88.739433 & -12.938778 & 7.73 & 15.78 & DR3 & 15.32 & 1 & 8 \\
27970 & 05550-1256 & Ac & 88.739433 & -12.938778 & 7.73 & 15.78 & DR3 & 1049.4 & 1 & 8 \\
28796 & 06048-4828 & A & 91.194499 & -48.458296 & 6.57 & 32.61 & DR3 & 2.51 & 1 & 5 \\
28796 & 06048-4828 & B & 91.195468 & -48.458666 & 7.69 & 32.47 & DR3 & \ldots & 0 & 5 \\
31089 & 06314+0749 & A & 97.857703 & 7.822383 & 8.28 & 14.40 & dyn & 212.9 & 1 & 7 \\
34212 & 07056-7116 & A & 106.404913 & -71.272168 & 7.66 & 16.55 & DR3 & 1246.7 & 1 & 8 \\
35261 & 07171-1202 & C & 109.275042 & -12.035929 & 9.44 & 18.98 & DR3 & 22.49 & 2 & 5 \\
35733 & 07223-3555 & B & 110.567716 & -35.918275 & 7.87 & 17.02 & DR3 & \ldots & 0 & 3 \\
35733 & 07223-3555 & A & 110.568225 & -35.916238 & 7.01 & 17.04 & DR3 & 4.63 & 2 & 3 \\
36160 & 07270-3419 & B & 111.740038 & -34.315499 & 8.19 & 20.71 & DR3 & \ldots & 0 & 9 \\
36165 & 07270-3419 & A & 111.744340 & -34.312096 & 7.03 & 20.32 & DR3 & 2280.0 & 1 & 9 \\
40523 & 08164-0314 & A & 124.110290 & -3.226089 & 7.25 & 12.84 & DR3 & 28.96 & 2 & 6 \\
41171 & 08240-1548 & A & 126.002870 & -15.797150 & 8.55 & 4.94 & DR3 & 25.42 & 2 & 6 \\
41171 & 08240-1548 & B & 126.002870 & -15.797150 & 8.55 & 4.94 & DR3 & 972.0 & 2 & 9 \\
44874 & 09086-2550 & A & 137.152546 & -25.839496 & 6.77 & 21.62 & DR3 & \ldots & 0 & - \\
45734 & 09194-7739 & A & 139.852773 & -77.643435 & 8.05 & 14.79 & DR3 & 7341.5 & 2 & 6 \\
45734 & 09194-7739 & B & 139.849891 & -77.645862 & 9.57 & 14.70 & DR3 & 0.56 & 1 & 6 \\
49336 & 10043-2823 & A & 151.074151 & -28.378782 & 7.29 & 10.67 & DR3 & 44.51 & 2 & 7 \\
49336 & 10043-2823 & B & 151.074151 & -28.378782 & 7.29 & 10.67 & DR3 & 1438.7 & 1 & 9 \\
51578 & 10321-7005 & A & 158.018467 & -70.084248 & 9.05 & 7.99 & DR3 & 2.19 & 1 & 6 \\
51578 & 10321-7005 & B & 158.014407 & -70.085932 & 9.33 & 7.97 & DR3 & \ldots & 0 & 6 \\
56282 & 11323-0025 & Aa & 173.077467 & -0.414750 & 8.09 & 16.28 & DR3 & 122.18 & 1 & 8 \\
56282 & 11323-0025 & Ac & 173.077467 & -0.414750 & 8.09 & 16.28 & DR3 & 3652.5 & 1 & 8 \\
57572 & 11480-6607 & A & 177.008890 & -66.114841 & 8.33 & 26.81 & DR3N & \ldots & 0 & 6 \\
57572 & 11480-6607 & B & 177.005754 & -66.112951 & 9.98 & 26.81 & DR3 & 36.40 & 2 & 6 \\
57860 & 11520-4357 & A & 178.004103 & -43.958411 & 8.40 & 12.09 & DR3 & 891.9 & 2 & 8 \\
59426 & 12114-1647 & A & 182.845486 & -16.790818 & 7.05 & 28.91 & DR3N & 211.6 & 2 & 6 \\
59426 & 12114-1647 & B & 182.844192 & -16.790187 & 8.69 & 29.35 & DR3 & \ldots & 0 & 6 \\
60845 & 12283-6146 & A & 187.070333 & -61.765703 & 6.82 & 19.97 & DR3 & 6.30 & 2 & 4 \\
60845 & 12283-6146 & B & 187.070333 & -61.765703 & 6.82 & 19.97 & DR3 & \ldots & 0 & 4 \\
62852 & 12530+1502 & A & 193.242303 & 15.031521 & 7.93 & 6.37 & DR3 & \ldots & 0 & - \\
62852 & 12530+1502 & B & 193.242112 & 15.029947 & 8.65 & 6.46 & DR3 & 13.93 & 2 & 6 \\
64478 & 13129-5949 & A & 198.232093 & -59.816601 & 6.16 & 23.20 & DR3 & 4.23 & 2 & 2 \\
64478 & 13129-5949 & B & 198.239820 & -59.822379 & 9.26 & 23.14 & DR3 & 0.24 & 2 & 2 \\
66438 & 13372-6142 & A & 204.301591 & -61.691863 & 5.63 & 27.89 & dyn & 8.07 & 2 & 6 \\
75663 & 15275-1058 & A & 231.876127 & -10.962551 & 8.14 & 7.73 & DR3N & 623.8 & 1 & 9 \\
75663 & 15275-1058 & B & 231.878159 & -10.964273 & 9.21 & 7.78 & DR3 & 22.87 & 2 & 4 \\
76400 & 15362-0623 & A & 234.053332 & -6.388504 & 7.90 & 15.28 & DR3 & 368.5 & 2 & 8 \\
76816 & 15410-1449 & A & 235.258012 & -14.823242 & 9.47 & 3.25 & DR3 & 6.95 & 2 & 4 \\
76816 & 15410-1449 & B & 235.256434 & -14.823258 & 9.74 & 3.32 & DR3 & 208.9 & 1 & 8 \\
78163 & 15577-3915 & A & 239.412807 & -39.249748 & 9.04 & 10.49 & DR3 & 21.82 & 2 & 4 \\
78163 & 15577-3915 & B & 239.414088 & -39.248463 & 10.30 & 10.65 & DR3N & 2082.5 & 1 & 9 \\
78416 & 16005-3605 & A & 240.130508 & -36.087939 & 8.65 & 9.35 & DR3 & 21.08 & 2 & 4 \\
78416 & 16005-3605 & B & 240.132709 & -36.088342 & 9.32 & 9.38 & DR3 & \ldots & 0 & 4 \\
79979 & 16195-3054 & B & 244.881427 & -30.901858 & 6.82 & 25.53 & DR3 & 1083.2 & 1 & 9 \\
79980 & 16195-3054 & A & 244.886410 & -30.906711 & 5.51 & 22.71 & DR3 & \ldots & 0 & 9 \\
80448 & 16253-4909 & A & 246.323273 & -49.147874 & 7.97 & 19.51 & DR3 & 2.27 & 2 & 4 \\
80448 & 16253-4909 & B & 246.324190 & -49.148091 & 8.23 & 19.10 & DR3 & \ldots & 0 & 4 \\
81394 & 16374-6133 & A & 249.350286 & -61.551652 & 8.54 & 5.02 & DR3 & \ldots & 0 & 8 \\
81395 & 16374-6133 & B & 249.350510 & -61.554676 & 9.22 & 5.04 & DR3N & 224.8 & 2 & 8 \\
82032 & 16454-7150 & A & 251.361957 & -71.839820 & 7.60 & 19.36 & DR3 & \ldots & 0 & - \\
82688 & 16541-0420 & A & 253.533931 & -4.340183 & 7.82 & 22.13 & DR3 & \ldots & 0 & - \\
84789 & 17199-1121 & A & 259.966441 & -11.351701 & 9.11 & 5.38 & DR3 & 2.28 & 2 & 4 \\
84789 & 17199-1121 & B & 259.965412 & -11.352876 & 9.89 & 5.41 & DR3 & \ldots & 0 & 4 \\
85342 & 17264-4837 & B & 261.593564 & -48.614901 & 7.05 & 21.83 & DR3 & \ldots & 0 & - \\
87813 & 17563-1549 & B & 269.081671 & -15.817828 & 9.13 & 12.49 & DR3N & 729.0 & 2 & 6 \\
88622 & 18056+0439 & A & 271.406056 & 4.657077 & 6.79 & 41.47 & DR3 & \ldots & 0 & - \\
88728 & 18068+0853 & A & 271.711656 & 8.875940 & 6.97 & 23.13 & DR3 & 1132.0 & 1 & 9 \\
92929 & 18560-2503 & A & 284.001731 & -25.045748 & 7.38 & 23.05 & DR3 & 312.3 & 1 & 5 \\
95106 & 19209-3303 & A & 290.223240 & -33.053035 & 8.16 & 20.25 & DR3 & \ldots & 0 & 3 \\
95110 & 19209-3303 & B & 290.226916 & -33.055194 & 9.72 & 20.86 & DR3 & 78.24 & 1 & 3 \\
95203 & 19221-2931 & A & 290.516098 & -29.523063 & 7.11 & 21.14 & DR3 & \ldots & 0 & - \\
98278 & 19583-5154 & B & 299.531614 & -51.905772 & 8.17 & 10.49 & DR3 & 236.2 & 2 & 8 \\
100420 & 20218-3654 & A & 305.448204 & -36.901513 & 8.31 & 5.66 & DR3N & 790.6 & 2 & 8 \\
100420 & 20218-3654 & B & 305.445819 & -36.900655 & 9.60 & 5.65 & DR3 & \ldots & 0 & 8 \\
101472 & 20339-2710 & A & 308.471892 & -27.171499 & 9.38 & 11.54 & DR3N & 354.9 & 2 & 6 \\
103735 & 21012-3511 & A & 315.305710 & -35.184168 & 7.66 & 22.10 & dyn & 4251 & 2 & 9 \\
103814 & 21022-4300 & A & 315.553079 & -43.002138 & 6.64 & 11.25 & DR3N & 1089 & 1 & 9 \\
104440 & 21094-7310 & A & 317.343607 & -73.172817 & 5.68 & 51.10 & dyn & 1934.7 & 1 & 9 \\
104835 & 21143-3835 & AB & 318.568071 & -38.577981 & 8.38 & 5.03 & DR3 & \ldots & 0 & 7 \\
104833 & 21143-3835 & C & 318.567886 & -38.580862 & 9.48 & 4.85 & DR3 & 11.34 & 1 & 7 \\
105441 & 21214-6655 & A & 320.352125 & -66.915967 & 8.77 & 31.39 & DR3 & 4.62 & 1 & 3 \\
105879 & 21266-4604 & A & 321.654889 & -46.067307 & 7.18 & 13.10 & DR3 & 2937 & 2 & 9 \\
105947 & 21274-0701 & A & 321.846906 & -7.015777 & 7.52 & 15.49 & DR3 & 8.73 & 1 & 3 \\
107731 & 21494-4759 & A & 327.352783 & -47.977590 & 8.64 & 15.82 & DR3N & 470.0 & 2 & 7 \\
109443 & 22104-5158 & A & 332.589393 & -51.960092 & 7.63 & 15.58 & DR3 & 970.9 & 1 & 9 \\
109951 & 22161-0705 & B & 334.027299 & -7.090689 & 8.74 & 15.90 & dyn & 111.1 & 1 & 3 \\
111598 & 22366-0034 & Aa & 339.140902 & -0.564739 & 8.38 & 6.68 & DR3 & 5.87 & 1 & 6 \\
111598 & 22366-0034 & Ac & 339.140902 & -0.564739 & 8.38 & 6.68 & DR3 & 271.2 & 1 & 6 \\
113597 & 23003-2609 & A & 345.116484 & -26.311881 & 9.65 & 31.48 & DR3 & 20.35 & 2 & 1 \\
115272 & 23208-5018 & A & 350.209036 & -50.306548 & 6.05 & 13.22 & DR3 & 3.43 & 1 & 5 \\
115269 & 23208-5018 & C & 350.205341 & -50.310466 & 8.86 & 13.33 & DR3 & \ldots & 0 & 5 \\
115552 & 23244+1429 & A & 351.099501 & 14.479084 & 7.60 & 6.38 & DR3 & 17.48 & 2 & 5 \\
117596 & 23509-7954 & A & 357.729881 & -79.898572 & 8.02 & 18.20 & DR3 & 4.37 & 2 & 5 \\
117666 & 23518-0637 & A & 357.945126 & -6.613069 & 8.73 & 13.30 & dyn & 780.9 & 1 & 9 \\
117666 & 23518-0637 & B & 357.945126 & -6.613069 & 8.73 & 13.30 & dyn & 253.8 & 1 & 9 \\
\enddata 
\tablenotetext{a}{Parallax references:
   DR3 -- Gaia Data Release 3;
   DR3N -- Gaia Data Release 3, NSS catalog;
   DR3* -- Gaia Data Release 3, other component of the system;
   HIP  -- Hipparcos; 
   dyn  -- dynamical (visual orbit and mass);
   orb  -- orbital (visual-spectroscopic orbit).
}
\tablenotetext{b}{SB orbit flags:
   0 -- no orbit;
   1 -- single-lined orbit;
   2 -- double-lined orbit.
}
\end{deluxetable*}

\begin{figure}
\epsscale{1.1}
\plotone{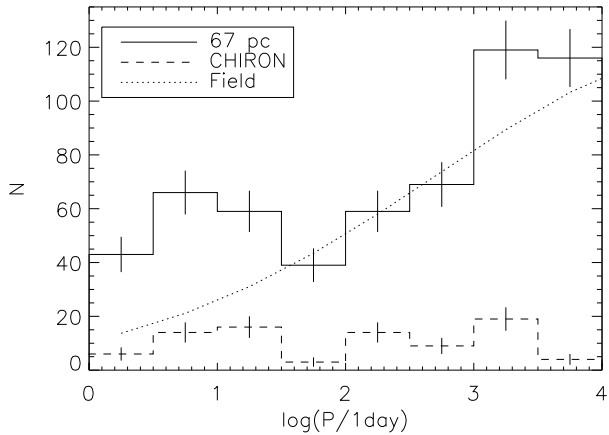}
\caption{Histogram  of  inner  periods   in  hierarchical  systems  of
  solar-type stars within  67\,pc (full line) and of the  subset of 90
  inner periods resulting from the  CHIRON program (dashed line).  The
  dotted line is a log-normal  period distribution of field solar-type
  binaries from \citet{R10} with arbitrary normalization.
\label{fig:chiper}
}
\end{figure}

The goal of  our survey was to  determine all (or most)  periods up to
1000 days. Naturally, some periods turned out to be longer than this
arbitrary threshold.  The distribution of  orbital periods  determined in
this survey is plotted in Figure~\ref{fig:chiper} in dashed line.  For
comparison, the distribution of periods  in all known inner subsystems
in hierarchies within 67\,pc with primary  star masses from 0.5 to 1.5
\msun  is plotted.   The dotted  line traces  the standard  log-normal
period distribution of field binaries with a median of $10^5$ days and
a logarithmic dispersion  of 2.28 dex \citep{R10}. At $P  > 100$ days,
both distributions  are similar,  showing an  increase of  orbits with
longer  periods. However,  inner subsystems  have a  strong excess  of
orbits with $P<30$ days compared to the canonical log-normal curve. In
other words, short-period binaries have  a strong preference to belong
to hierarchical  systems.  This phenomenon is  further discussed below
in Section~\ref{sec:pin}.  Neither the MSC  nor the CHIRON samples are
complete, but, owing to the  extended monitoring, periods shorter than
$10^3$ days should  be represented uniformly. So, the  excess of short
periods  and  the local  minimum  in  the  30--100  days bin  are  real
features rather than selection effects.

\section{Comparison between CHIRON and Gaia NSS Orbits}
\label{sec:dr3}

\begin{figure*}
\epsscale{1.0}
\plotone{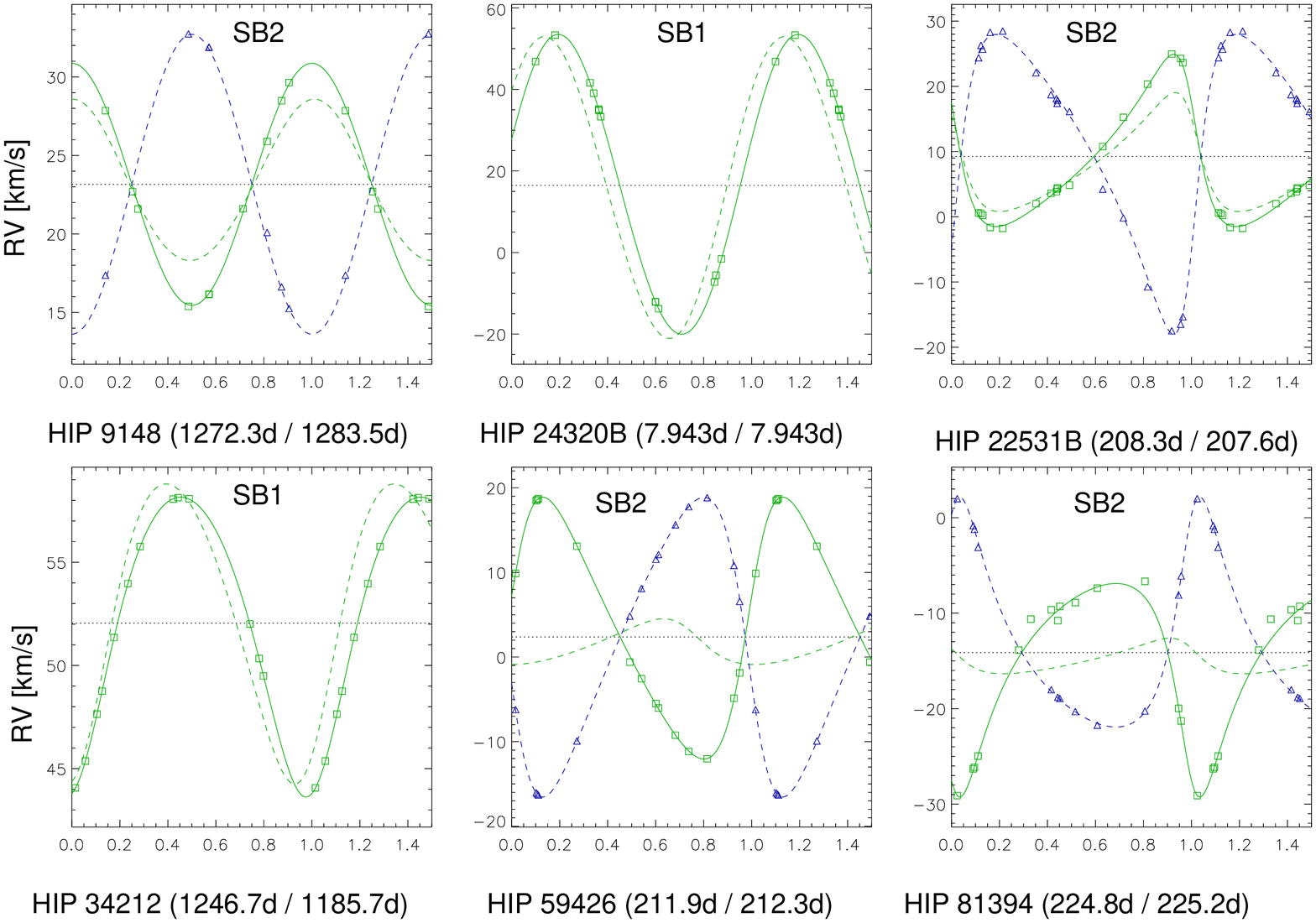}
\caption{Comparison between  six CHIRON and Gaia  spectroscopic orbits.
   All plots show RVs of the primary (solid green
  line and  squares) and  secondary (dashed  blue line  and triangles)
  components measured by  CHIRON vs.  orbital phase.   The Gaia orbits
  are depicted by dashed green lines.  The CHIRON and Gaia periods are
  indicated under each plot.
\label{fig:gaia}
}
\end{figure*}

\begin{deluxetable}{ccc c c  c }
\tabletypesize{\scriptsize}     
\tablecaption{Comparison between CHIRON and Gaia NSS\tablenotemark{b} Orbits
\label{tab:chinss} }  
\tablewidth{0pt}                                   
\tablehead{                                                                     
\colhead{HIP} & 
\colhead{Comp.} &
\colhead{SB1/2} &
\colhead{$P_{\rm CHI}$} & 
\colhead{NSS\tablenotemark{a}} &
\colhead{$P_{\rm NSS}$} \\
& & & 
\colhead{(days)} &      
\colhead{sol.} &
\colhead{(days)}
}
\startdata
  1103 & A &\ldots  &\ldots     & SB1  & 1343.0 \\  %
  6873 & B & SB2 & 115.6  & SB2  & 112.4 \\  
  7852 & A & SB1 & 1177.5 & SB1  & 1184.0 \\ 
  9148 & A & SB2 & 1272.3 & ASB1 &  1283.5\tablenotemark{b} \\ 
 16853 & A &\ldots  &\ldots     & ASB1 &  204.1 \\ 
 21079 & A & SB1 &  217.0 & SB1  & 2.71\tablenotemark{c} \\ 
 22531 & A & SB1 & 1003   & AORB & 1000.8 \\ 
 22534 & B & SB2 & 208.3  & ASB1 & 207.6\tablenotemark{b} \\ 
 24320 & A & SB1 & 1430.3 & SB1  & 1042.9 \\ 
 24320 & B & SB1 & 7.943 & SB1  & 7.943 \\ 
 34212 & A & SB1 & 1246.7 & SB1  & 1185.6 \\ 
 35733 & A & SB2 & 4.63   & SB2  & 4.63 \\  
 51578 & A & SB1 & 2.19   & SB1  & 2.19 \\ 
 56282 & A & SB1 & 121.1  & SB1  & 125.3 \\ 
 57572 & A &\ldots  &\ldots     & ASB1 & 169.6 \\ 
 59426 & A & SB1 & 211.6  & ASB1 & 212.3\tablenotemark{b} \\ 
 64478 & A & SB2 & 4.23   & SB2  & 4.23 \\ 
 75663 & A & SB1 & 623.8  & ASB1 & 626.7\tablenotemark{b} \\ 
 78163 & B & SB1 & 2082.5 & AORB & 1532.4 \\  
 81395 & B & SB2 & 224.8  & ASB1 & 225.2\tablenotemark{b} \\ 
 84789 & A & SB2 & 2.28   & SB2  & 2.28 \\ 
 88728 & A & SB1 & 1132.0 & SB1  & 1267.8\tablenotemark{b}\\
100420 & A & SB2 & 790.6 & AORB  & 805.4 \\ 
101472 & A & SB2 & 354.9 & AORB  & 354.0 \\ 
103814 & A & SB1 & 1089.0 & ASB1 & 1119.7 \\ 
104833 & C & SB1 & 11.34  & SB1  & 11.34 \\  
105441 & A & SB1 & 4.62   & SB1  & 4.62 \\ 
105441 & B &\ldots  & \ldots    & ASB1 & 549.5 \\  %
107731 & A & SB2 & 469.9  & ASB1 & 469.3\tablenotemark{b} \\ 
109443 & A & SB1 & 970.9  & SB1 & 989.8 \\ 
115552 & A & SB2 & 17.48  & SB2 & 17.48 \\ 
\enddata
\tablenotetext{a}{NSS solutions:
SB1 --- single-lined spectroscopic;
SB2 --- double-lined spectroscopic;
ASB1 --- astrometric and single-lined spectroscopic;
AORB --- astrometric.
}
\tablenotetext{b}{Reduced RV amplitude}
\tablenotetext{c}{Wrong period}
\end{deluxetable}

The  number of  spectroscopic  orbits in  GDR3/NSS ($ 1.8\times  10^5$)
overwhelms  all  spectroscopic  orbits  determined to  date  from  the
ground:   on   March   24,  2020,   the   SB9   catalog\footnote{{\url
    https://sb9.astro.ulb.ac.be/}} \citep{Pourbaix2004} contained only
4004 systems, with 2/3 of those  on the northern sky.  Gaia determined
orbits      by       an      impersonal       automated      procedure
\citep{Arenou2022,Pourbaix2022}.  The duration of the GDR3 mission (34
months)  and  the observing  cadence  set  by  the Gaia  scanning  law
naturally restrict the range of  accessible orbits.  Most Gaia periods
are under  1000 days, and  orbital periods close  to one year  and its
harmonics are  underrepresented.  The distribution of  Gaia orbits on
the sky  is nonuniform and clearly  shows an imprint of  the scanning
law.  Candidates for  Gaia orbit determination went  through a vetting
procedure (visual binaries with  close separations were rejected), and
the orbits  were checked by various  filters.  Documentation available
on the  Gaia web site  \citep{Pourbaix2022} describes the  vetting and
quality control.   Comparison with known spectroscopic  orbits  in
  the cited  document indicated a ``recovery  rate'' (correctness) of
Gaia  spectroscopic  orbits between  0.7  and  0.9, depending  on  the
comparison sample and criteria used.

The CHIRON  sample presented  here was matched  by coordinates  to the
Gaia catalogs  of single-  and double-lined spectroscopic  orbits (SB1
and  SB2, respectively)  and other  NSS solutions.   Coordinate search
reveals 31 Gaia orbits among CHIRON targets; four of those have no CHIRON
orbits owing  to the small number  of observations, for five  targets NSS
contains      only      astrometric     orbits      with      matching
periods. Table~\ref{tab:chinss} compares the CHIRON and NSS orbits for
31 common targets.  The  NSS solution codes are obvious (SB1  and SB2 for
single-    and   double-lined    spectroscopic   orbits,    ASB1   for
spectro-astrometric orbits,  and AORB for purely  astrometric orbits).
One NSS  orbit (HIP 21079A) is  false owing to the  wrongly determined
period of 2.71  days (the true period is 217  days).  Seven NSS orbits
have reduced  RV amplitudes, either  because of blending with  lines of
the secondary or because of an inaccurate  shape of the NSS RV curve. The
remaining  14 SB  orbits  in  common between  CHIRON  and  NSS are  in
reasonably good mutual agreement.  Note, however, that NSS missed some
subsystems,  either inner  (1.56 days  in  HIP 22531A)  or outer  with
periods exceeding the  GDR3 mission duration (in  HIP 56282A, 64478A).
For  HIP~36165  with  $P=2300$  days,  Gaia  detected  so  far  only
the acceleration and the RV trend.  Overall, only a third of the CHIRON sample
has  spectroscopic  or astrometric  orbits  in  the NSS.   The  CHIRON
targets are  bright (well above the  Gaia RV threshold of  13 mag), and
most orbital periods are under 1000 days.  So, despite the large total
number of orbits, the NSS orbit catalog is still very incomplete.

Figure~\ref{fig:gaia} compares  CHIRON and  Gaia orbits of  six common
stars  where  the  periods  match  approximately.   Four  systems  are
double-lined in CHIRON, but single-lined  in Gaia.  When the secondary
lines are much  fainter than those of the primary  (HIP 9148, 22531B),
Gaia gives  a reasonable match  for the  primary star with  a slightly
reduced RV amplitude.  For  pairs with comparable-mass components (HIP
59426,   81394),    the   Gaia   RV   amplitudes    are   dramatically
underestimated; moreover, the  shape and phase of the  Gaia RV curves
are incorrect.  For two single-lined binaries (HIP 24320B, 34212), the
CHIRON and  Gaia orbits  are similar  (small phase  shifts are  due to
inaccurate Gaia periods). Interestingly, a  comparison of the NSS with two
large ground-based RV surveys indicated that only about a half of Gaia
SB1 orbits could be validated \citep{Bashi2022}.

The availability of Gaia orbits is  most welcome. However, it does not
make the  CHIRON survey  obsolete, quite  to the  contrary. Presently,
Gaia provides orbits for only a  small fraction of inner subsystems in
nearby hierarchies, and some of those orbits are questionable even for
the relatively  simple binaries shown in  Figure~\ref{fig:gaia}.  More
complex   and   more   interesting    systems   (e.g.    triple-   and
quadruple-lined)  can  be discovered  and  studied  only by  dedicated
ground-based programs like this one.  Systematic underestimation of RV
amplitudes by Gaia  due to line blending leads  to the underestimated
mass ratios,  so the use of  Gaia orbits for a statistical  study of the
mass ratio distribution is not recommended.

\section{Radial Velocities of Wide Pairs}
\label{sec:wide}

\begin{deluxetable}{r l  cc cc}    
\tabletypesize{\scriptsize}     
\tablecaption{RVs of Stars in Wide Pairs (Fragment)
\label{tab:rvtable}          }
\tablewidth{0pt}                                   
\tablehead{                                                                     
\colhead{HIP} & 
\colhead{Comp.} & 
\colhead{JD} & 
\colhead{RV} & 
\colhead{$a$} & 
\colhead{$\sigma$}  \\
&  &
\colhead{$-2\,400\,000$} &
\colhead{\kms} &
 &
\colhead{\kms} 
}
\startdata
  1103& A & 57276.7918 &   2.785 & 0.056 & 20.455 \\
  1103& A & 57299.7225 &   2.729 & 0.057 & 20.539 \\
  1103& A & 57333.5715 &   2.814 & 0.057 & 20.569 \\
  1103& A & 57983.7853 &   5.967 & 0.057 & 20.178 \\
  2713& A & 57986.7734 &   8.987 & 0.353 &  4.345 \\
  2715& C & 57986.7745 &   5.899 & 0.307 &  4.730 \\
  5896& B & 57985.7965 &   8.228 & 0.457 &  4.465 \\
  5896& A & 57985.7954 &   8.042 & 0.036 & 31.149 \\
  6712& A & 57985.8439 & -23.365 & 0.471 &  3.844 
\enddata 
\end{deluxetable}

Wide (resolved)  pairs were probed  for the presence of  subsystems by
measuring  RVs   of  each   component  and  looking   for  substantial
differences  \citep{Tok2015}.  This  work  has been  continued in  the
following years and its complementary  results are reported here. Some
wide  pairs contain  known visual  subsystems with  long periods,  and
their RVs change  on a time scale of decades.   The RV measurements of
wide pairs with CHIRON are published in Table~\ref{tab:rvtable}, to be
used  in the  future for  determination of  long-period orbits.   This
table also contains  previously unpublished RVs of stars from the
main program where insufficient number  of observations does not allow
for orbit calculation  or the  orbits are known  from other  sources.  For
example, the orbital period of HIP~1103A, 1343 days, is now determined
by Gaia, and its continued  monitoring with CHIRON makes little sense.
Columns of the table identify stars  by their HIP number and component
letter.  Then  follow  Julian date,  RV, amplitude $a$, and  rms
width $\sigma$ of  the cross-correlation dip.  The  dip parameters are
helpful in evaluating the RV errors (wide and shallow dips give larger
errors), while  variable dips indicate blending  of several components
with  variable RVs.  Table~\ref{tab:rvtable}  contains 162  RVs of  91
distinct targets.

Five stars  in Table~\ref{tab:rvtable}  have multi-component  dips and
deserve  individual   comments.   Triple  lines  in   HIP~49442A  were
discovered  here  and  indicate  a  spectroscopic  subsystem  in  this
0\farcs18 visual pair; the RV of  the B component, at 4\farcs4 from A,
was also measured.  The double  dip of HIP~61465B signals a subsystem,
in agreement with the astrometric signature of an unresolved binary in
Gaia;  its counterpart  HIP~61466A  at 27\farcs5  may  also contain  a
subsystem. HIP~78662C (a $V=8$ mag  star at 11\arcsec ~from the bright
young visual binary HIP~78662AB) may have  a double dip, but its large
width and small amplitude render this discovery uncertain.  The modest
RV difference between the dip  components in HIP~79588AB may be caused
by  motion in  the 34  yr visual  orbit with  a large  eccentricity of
0.8. HIP~111391AB is also a visual binary  with a period of 198 yr and
an  eccentric orbit  which likely  causes  the small  and constant  RV
difference between the two dip components.

\section{Statistics of Inner Subsystems}
\label{sec:stat}

\subsection{Period-Eccentricity Diagram}
\label{sec:pe}

\begin{figure}
\epsscale{1.1} 
\plotone{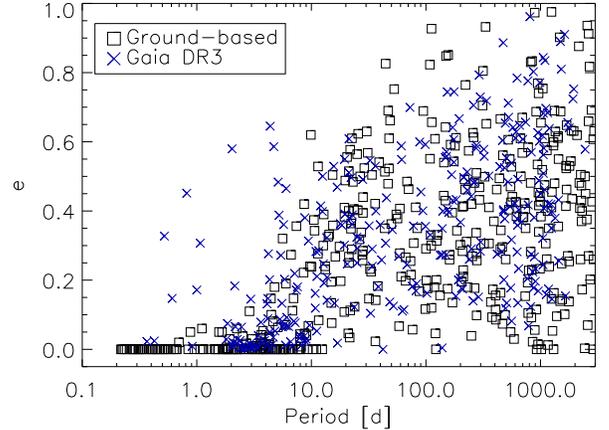}
\caption{Periods and eccentricities of   743 inner subsystems with
  solar-type components  found in the MSC. Squares correspond to the
  ground-based  orbits,  blue crosses are orbits from Gaia NSS. 
\label{fig:pe}
}
\end{figure}

Patient accumulation of data on nearby solar-type hierarchies improves
completeness of  their sample.  The  multi-year CHIRON survey  and the
Gaia  DR3  greatly  reduce  the   historic  bias  in  favor  of  short
periods. Although not all inner orbits with $P < 3000$ days are known,
the observational ``window''  is more uniform, and the  number of known
orbits is larger than a decade ago. So, a fresh look at the statistics
in the short-period regime is warranted using the up-to-date MSC.

I selected from the MSC inner  subsystems with primary masses from 0.7
to  1.5  \msun  (solar-type),  known  periods  under  3000  days,  and
distances within 100\,pc --- a  total of  743 cases with spectroscopic,
astrometric,  or  visual  inner  orbits (455  ground-based,  288  from
Gaia).  References  to the orbits can  be found in the  MSC and in
  SB9  \citep{Pourbaix2004}.    The  median  primary  mass   is  0.99
\msun. If the distance limit is reduced to 67\,pc, the balance between
ground-based (325)  and Gaia (53)  orbits shifts further,  showing 
improved completeness of the ground-based data for nearby stars.

Figure~\ref{fig:pe} plots  the period-eccentricity relation  for inner
subsystems within 100\,pc.  The upper  envelope of the points outlines
the tidal circularization: most orbits with $P < 10$ days are circular
\citep{Meibom2005}.   Several  crosses  at $P<10$  days,  outside  the
envelope,  are  Gaia  orbits  with spurious  periods  that  should  be
ignored. The large  number of orbits reveals  an interesting structure
in this  diagram, such  as two concentrations  of  eccentric
orbits with  periods of 10--30 days  and with $P>100$ days,  while the
number of  orbits in  the intermediate  30--100 days  interval appears
smaller and these orbits seem to have smaller eccentricities. Circular
orbits reappear again at $P > 1000$ days.

Period-eccentricity diagrams  for spectroscopic binaries can  be found
in  many  papers  \citep[e.g.][]{R10,Triaud2017,PW2020,Torres2021};  a
circularization  period  $P_{\rm  circ}$  between 7  and  10  days  is
inferred  from  these plots.   A  few  eccentric orbits  with  periods
shorter than  $P_{\rm circ}$  seen in such  diagrams are  explained by
inefficient tides  in stars  with primary masses  above 1.3  \msun and
small secondaries  \citep{Triaud2017} or by the  influence of tertiary
companions     \citep{Mazeh1990};     all    close     binaries     in
Figure~\ref{fig:pe} are inner subsystems  within multiple stars.  Most
short-period  eccentric  orbits  derived by  \citet{PW2020}  ``in  the
regime of sparse, noisy, and  poorly sampled multi-epoch data'' likely
are spurious.

All binaries  with periods less than  $\sim$10 yr were formed  by some
kind of migration,  although the migration mechanisms  are still under
debate \citep{Moe2018}.  The  $P - e$ diagram may  be instructive from
this perspectuve. It shows that some inner pairs with $ P > 1000$ days
could be formed with quasi-circular  orbits (many visual binaries with
nearly circular orbits and periods on  the order of a decade are known
as  well).  However,  further shortening  of  the period  seems to  be
associated with an eccentricity growth, given that small eccentricites
are rare  at periods of  $\sim 10^2$ days.  The  mechanism responsible
for  migration in  this regime  should be  associated with  a loss  of
angular momentum,  e.g. via  interaction with  a circumbinary  disk or
with outer companions.  The growing eccentricity and shortening period
eventually  bring the  pair into  a regime  of tidal  circularization,
where separation at  periastron is a few  stellar radii. Concentration
of  points  at  $ P  \sim  20$  days  and  $e  = 0.4  \ldots  0.6$  in
Figure~\ref{fig:pe} corresponds  to inner pairs that  have reached the
tidal  regime  and  apparently  start slow  evolution  toward  shorter
periods and circular orbits.  Interestingly, \citet{DOrazio2021} found
by hydrodynamical  simulation that an  eccentric binary in  a coplanar
prograde  disc   evolves  to   shorter  periods,   while  eccentritity
fluctuates around the $e=0.4$ attractor; in contrast, a binary with $e
< 0.1$ does not migrate and its orbit remains circular.

\subsection{Period Distribution of Inner Subsystems}
\label{sec:pin}

The histogram of inner periods in solar-type hierarchies is plotted in
Figure~\ref{fig:chiper}. It shows a  marked difference with the period
distribution  of all  field  binaries, namely  the  excess of  periods
shorter  than  30  days  in inner  subsystems.   Preference  of  close
binaries to be  members of hierarchical systems  is firmly established
by  prior   work.   For   example,  \citet{Tok2006}   determined  that
$\sim$80\%  of spectroscopic  binaries with  $P<7$ days  have tertiary
companions.   The  statistical   model  of  hierarchical  multiplicity
developed  in  \citep{FG67b}  matches   the  data  quite  well  (after
accounting  for the  selection),  but underpredicts  the fraction  of
inner subsystems with $P < 10$ days by a factor of two, because it does
not account  for correlation  between close binaries  and higher-order
multiplicity.  Recently  \citet{Hwang2020} found that  the occurrence
rate of wide physical companions  to eclipsing  binaries is  $\sim$3 times
higher than for typical field stars (14.1\% vs. 4.5\%).

The  origin  of  the  observed relation  between  close  binaries  and
hierarchies  is  still  under  debate.   The  reader  is  referred  to
\citet{Moe2018} for  a detailed analysis and  an up-to-date population
synthesis. The originally proposed  mechanism of Lidov-Kozai cycles in
misaligned  triples  acting in  combination  with  tidal friction  can
account only  for a  minor fraction  of the  close subsystems,  and its
predictions disagree with reality.  Specifically, the predicted excess
of inner  periods just below the  tidal cutoff at $P<10$  days and the
concentration  of mutual  inclinations  near 40\degr  ~are not  seen.
Instead,  there  is no  discontinuity  in  the distribution  of  inner
periods at $P \sim  10$ days, but their numbers drop at  $P > 30$ days
(Figure~\ref{fig:chiper}).  \citet{Moe2018} argue  that the main agent
that shrinks inner periods should  be associated with the accretion of
gas during mass  assembly. The observed frequency of  inner twins with
mass ratio $q >  0.95$ formed by accretion also drops  sharply at $P >
30$ days  \citep[see Figure~4 in][]{Mult2021}. However,  the key issue
of relating accretion to the  presence of tertiary companions is still
unsettled.  Further  discussion of this  topic is beyond the  scope of
this paper.  The point here is  to illustrate how new homogeneous data
on hierarchies contribute to the study of their formation mechanisms.

\section{Summary}
\label{sec:sum}

Multi-year spectroscopic monitoring of  inner subsystems in solar-type
hierarchies  has  been undertaken  to  elucidate  distributions  of
periods, eccentricities,  and mass  ratios.  The targets  were derived
from the sample  of solar-type stars within 67\,pc  with unknown inner
periods  and   enlarged  by   additional,  more   distant  hierarchies.
Spectroscopic (and visual) orbits based  on these data are reported in
10 papers.  The main results are as follows.

\begin{itemize}
\item
A  total of  102  spectroscopic  orbits with  periods  ranging from  
fraction of  a day to several  years were determined. The  coverage is
reasonably complete up to $P \sim 1000$ days.

\item
Monitoring with  CHIRON revealed  new, additional  subsystems: several
presumed triples in fact are quadruples of 2+2 or 3+1 hierarchy.

\item
The Gaia  NSS provides orbits for  only a third of  the CHIRON sample,
and a substantial  fraction of Gaia orbits in common  with CHIRON have
reduced RV amplitudes or other problems.

\item
The distribution  of inner periods  based on this  survey, literature,
and  Gaia   (Figure~\ref{fig:chiper}),  differs  from   the  canonical
log-normal period distribution  in the field. Inner  subsystems have a
strong excess of periods shorter  than 30 days.  The logarithmic inner
period distribution has a local minimum in the 30--100 days bin.

\item
The  period-eccentricity   diagram  of  inner   solar-type  subsystems
(Figure~\ref{fig:pe}) shows an interesting structure. Statistical data
on periods, eccentricities, and mass ratios in these hierarchies will
help in the development and verification of their formation models.


\end{itemize}

In  a broader  context, this  study fits  into the  vast landscape  of
observational   characterization   of  stellar   hierarchies.    Large
photometric surveys  designed for  the study of  transiting exoplanets
have opened a new window on  unusual and rare compact hierarchies like
triply  eclipsing  planar  worlds  \citep[e.g.][]{Rappaport2022}  that
challenge  current formation  theories.  Gaia  has revolutionized  the
census of solar neighborhood by revealing  wide pairs of stars with an
unprecedented  completeness and  their  connection  to close  binaries
\citep[e.g.][]{Hwang2020,Hwang2022b}. New data highlight the diversity
of stellar hierarchies \citep{Mult2021}  and provide insights on their
formation \citep{Offner2022}.




\begin{acknowledgments}

The research  was funded  by the  NSF's NOIRLab.   This work  used the
SIMBAD   service  operated   by   Centre   des  Donn\'ees   Stellaires
(Strasbourg, France),  bibliographic references from  the Astrophysics
Data System  maintained by  SAO/NASA, and  the Washington  Double Star
Catalog maintained at  USNO.  This work has made use  of data from the
European     Space     Agency     (ESA)     mission        Gaia
(\url{https://www.cosmos.esa.int/gaia}), processed  by the  Gaia
Data      Processing      and     Analysis      Consortium      (DPAC,
\url{https://www.cosmos.esa.int/web/gaia/dpac/consortium}).    Funding
for the DPAC has been provided by national institutions, in particular
the  institutions   participating  in  the     Gaia  Multilateral
Agreement.  

\end{acknowledgments} 

\facility{CTIO:1.5m, Gaia}







\end{document}